\documentclass[aps,amsmath,notitlepage,twocolumn,amssymb,prl,showpacs,longbibliography]{revtex4-1}

\usepackage{graphicx}% Include figure files
\usepackage{dcolumn}% Align table columns on decimal point
\usepackage{bm}% bold math
\usepackage[colorlinks=true,citecolor=red,urlcolor=blue]{hyperref}
\usepackage{wasysym}
\usepackage{stmaryrd}
\usepackage{verbatim}
\usepackage{subfigure}
\usepackage{amsmath}

\begin{document}
\title{Rare-earth chalcogenides: A large family of triangular lattice spin liquid candidates}

\author{Weiwei Liu$^{1,2}$}
\thanks{These authors contributed equally to this work.}
\author{Zheng Zhang$^{1,2}$}
\thanks{These authors contributed equally to this work.}
\author{Jianting Ji$^{1}$}
\thanks{These authors contributed equally to this work.}
\author{Yixuan Liu$^{2}$}
\author{Jianshu Li$^{1,2}$}
\author{Xiaoqun Wang$^{3}$}
\author{Hechang Lei$^{2}$}
\email{hlei@ruc.edu.cn}
\author{Gang Chen$^{4}$}
\email{chggst@gmail.com}
\author{Qingming Zhang$^{1,5}$}
\email{qmzhang@iphy.ac.cn}
\affiliation{$^{1}$National Laboratory for Condensed Matter Physics and Institute of Physics, Chinese Academy of Sciences, Beijing 100190, China}
\affiliation{$^{2}$Department of Physics and Beijing Key Laboratory of Opto-electronic Functional Materials \& Micro-nano Devices, Renmin University of China, Beijing 100872, China}
\affiliation{$^{3}$Department of Physics and Astronomy, Shanghai Jiao Tong University, Shanghai 200240, China}
\affiliation{$^{4}$State Key Laboratory of Surface Physics and Department of Physics, 
Fudan University, Shanghai 200433, China}
\affiliation{$^{5}$School of Physical Science and Technology, Lanzhou University, Lanzhou 730000, China}

\date{\today}

\begin{abstract}
Frustrated quantum magnets are expected to host many exotic quantum spin states 
like quantum spin liquid (QSL), and have attracted numerous interest in modern 
condensed matter physics. The discovery of the triangular lattice spin liquid 
candidate YbMgGaO$_4$ stimulated an increasing attention on the rare-earth-based 
frustrated magnets with strong spin-orbit coupling. Here we report the synthesis 
and characterization of a large family of rare-earth chalcogenides AReCh$_2$ 
(A = alkali or monovalent ions, Re = rare earth, Ch = O, S, Se). The family compounds 
share the same structure (R$\bar{3}$m) as YbMgGaO$_4$, and antiferromagnetically 
coupled rare-earth ions form perfect triangular layers that are well separated 
along the $c$-axis. Specific heat and magnetic susceptibility measurements on NaYbO$_2$, 
NaYbS$_2$ and NaYbSe$_2$ single crystals and polycrystals, reveal no structural 
or magnetic transition down to 50mK. The family, having the simplest structure 
and chemical formula among the known QSL candidates, removes the issue on possible 
exchange disorders in YbMgGaO$_4$. More excitingly, the rich diversity of the family 
members allows tunable charge gaps, variable exchange coupling, and many other 
advantages. This makes the family an ideal platform for fundamental research of 
QSLs and its promising applications.
\end{abstract}


\maketitle

\emph{Introduction.}---The concept of quantum spin liquids (QSLs) was originally 
proposed by P. W. Anderson theoretically over 40 years ago~\cite{pwanderson}. 
It describes a highly entangled quantum state for spin degrees of freedom 
and was initially constructed with a superposition of spin singlets on the 
triangular antiferromagnet, so-called resonating-valence-bond state~\cite{pwanderson}. 
Later on, the possible connection between QSLs and high-temperature  
superconductivity was theoretically established through doping a QSL 
Mott insulator~\cite{RevModPhys.78.17}. Although the underlying mechanism 
for the high-temperature superconductivity has not yet come into a 
consensus, our understanding of QSLs has greatly improved, both from  
exactly solvable models~\cite{Kitaev,Wen2007} and several 
classification schemes~\cite{Wen2007,PhysRevB.87.104406}. 
On the experimental side, various frustrated magnetic materials, particularly 
the triangular-lattice-based antiferromagnets, were considered to be the most 
promising systems to realize QSLs~\cite{Balents2010}. So far, a number 
of compounds have been reported to host QSLs. Among them, the well-known 
ones include herbertsmithite and its derived 
compounds~\cite{kagome,Han2012,kagome_NMR,vesignieite,voborthite,volborthite2,li14,zorko08}, 
and triangular organics~\cite{dmit,organics2,kappaET,organictherm,organics1}. 
The magnetic ions in most of these compounds are $3d$ transition metal 
ions Cu$^{2+}$ with ${S=1/2}$, which may be crucial to enhance quantum 
fluctuations.

\begin{figure*}[t]
\includegraphics[width=16cm]{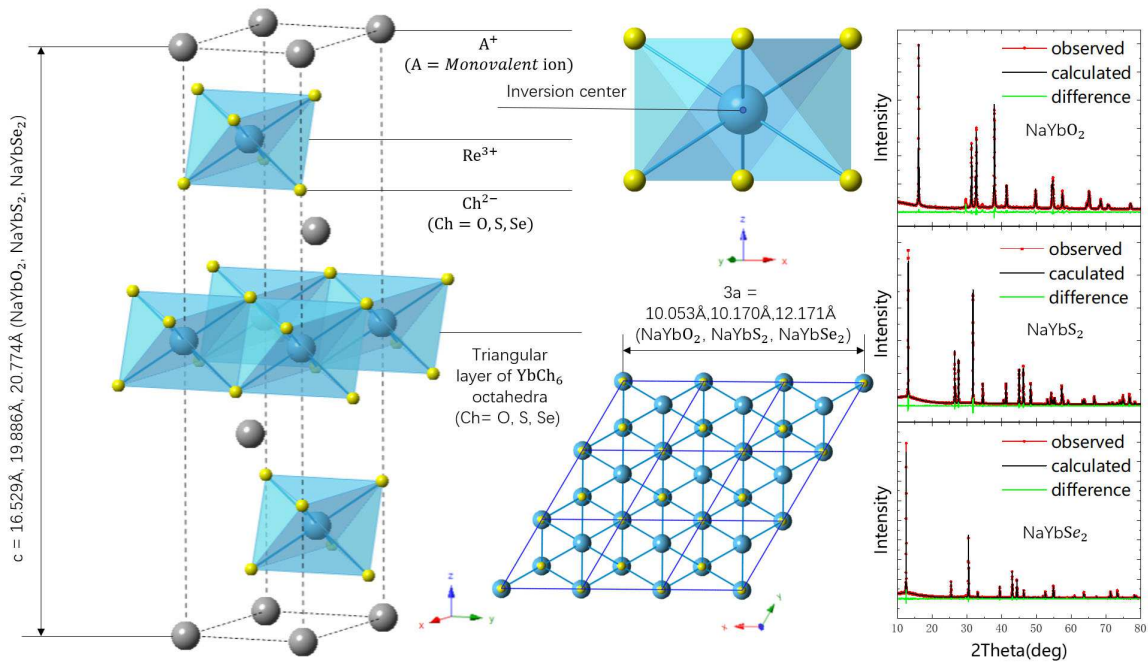}
\caption{(Color online.) The general crystal structure of rare-earth chalcogenides 
and the powder diffraction patterns and Rietveld refinements for NaYbCh$_2$ 
(Ch = O, S, Se).}
\label{fig1}
\end{figure*}

Quite recently, frustrated materials with magnetic rare-earth ions are proposed to 
be promising QSL candidates~\cite{RevModPhys.82.53}. 
These include the well-known pyrochlore ice materials~\cite{QSI1,YbTiO2,PrSnO,Ross2011,TbTiO,Onoda2010,YbTiO2012,YbTiO2012b,PhysRevLett.112.167203,PhysRevB.86.104412},
the kagome magnet~\cite{PhysRevLett.116.157201,Shulei}, 
and the triangular lattice magnets~\cite{YueshengSR,YueshengPRL1,PhysRevLett.117.097201,YMGOYao,nphys2017,PhysRevLett.117.267202,PhysRevX.8.031001,PhysRevLett.118.107202,YueshengTMGO,nnrvb,Wen,arXiv1708.06655,Yaodong1,GCoctu,GCnonK}. 
The local degree of freedom for the rare-earth ions that 
contain an odd number of $4f$ electrons (excluding Gd$^{3+}$), 
is a Kramers doublet and can be mapped to an effective spin ${S=1/2}$
degree of freedom. This effective-spin local moment is protected by time reversal symmetry 
and the point group symmetry. In many cases the non-Kramers rare-earth 
ions can be taken as effective spin ${S=1/2}$ 
local moments at low temperatures, though lacking the protection from time 
reversal symmetry~\cite{Curnoe2008,PhysRevB.83.094411,Onoda2010,PhysRevB.86.104412,GCnonK}.
The spin-orbit-entangled nature of the rare-earth local moments often brings 
highly anisotropic spin models that have never been constructed and 
studied before~\cite{Curnoe2008,PhysRevB.83.094411,Onoda2010,Yaodong1,GCoctu,GCnonK}. 
Thus, the rare-earth-based magnets play an important role in the 
exploration of novel spin models and the exotic magnetic states on various lattices. 
Indeed, QSL behaviors and multipolar phases have been proposed for various 
rare-earth compounds~\cite{QSI1,YbTiO2,PrSnO,Ross2011,TbTiO,Onoda2010,
YbTiO2012,YbTiO2012b,PhysRevLett.112.167203,PhysRevB.86.104412,Shulei,GCoctu,GCnonK}.

The recent discovery of the triangular lattice magnet YbMgGaO$_4$ has 
invoked a further interest in the search of spin liquids with strong
spin-orbit coupling~\cite{YueshengSR,YueshengPRL1,Yaodong1,PhysRevB.96.054445,
PhysRevB.97.184413,PhysRevB.97.125105,arXiv1708.06655,PhysRevB.96.075105,PhysRevLett.119.157201}. 
The compound has a space group symmetry of R$\bar{3}$m, 
and the Yb$^{3+}$ ions form a flat and perfect triangular lattice~\cite{YueshengSR,YueshengPRL1}
The availability of high-quality single crystals allows extensive and 
careful studies of magnetic properties using neutron 
scattering~\cite{YMGOYao,nphys2017,PhysRevLett.117.267202,PhysRevX.8.031001}, 
muon spin relaxation (muSR)~\cite{PhysRevLett.117.097201}, electron 
spin resonance (ESR)~\cite{YueshengPRL1} etc. These studies point to a possible gapless 
U(1) QSL ground state~\cite{YueshengSR,YueshengPRL1,PhysRevLett.117.097201,YMGOYao,
nphys2017,PhysRevX.8.031001,PhysRevLett.118.107202,PhysRevB.96.075105,PhysRevB.97.125105}. 
On the other hand, some experiments and 
theoretical arguments raised the issue on Ga/Mg disorder, which was 
suggested to be responsible for the disordered spin state and/or
QSL stability~\cite{YueshengSR,nphys2017,PhysRevLett.119.157201,PhysRevLett.120.207203,
PhysRevX.8.031001,PhysRevX.8.031028}. 
The small exchange coupling allows an easy tunability of the spin state 
with a laboratory magnetic field~\cite{YueshengSR,nphys2017,arXiv1708.06655,PhysRevX.8.031001}. 
Meanwhile, it also requires that most experiments must be carried out 
at ultralow temperatures. In some cases this could be an obstacle 
for in-depth studies and possible applications.

As mentioned above, there is a long list of rare-earth magnets~\cite{RevModPhys.82.53,GCnonK,Yaodong1}. 
Then the question is if one can find out some interesting compounds or 
systems with larger exchange couplings and without disorder. 
This is the purpose of this work. We systematically synthesized 
the rare-earth chalcogenides AReCh$_2$ (A = alkali or monovalent ions, 
Re = rare earth, Ch = O, S, Se) with a delafossite structure. We carried out 
the structural and thermodynamic characterizations of these compounds. 
The compounds have a high symmetry of R$\bar{3}$m and perfect spin 
triangular layers. The magnetic measurements indicate that spins 
are antiferromagnetically coupled in all the compounds with a range of 
Curie-Weiss temperatures. For the representative NaYbCh$_2$ (Ch = O, S, Se) 
samples, no magnetic ordering or transition is observed in the specific 
heat and susceptibility measurements down to 50mK. Thus, this is a large 
family of QSL candidates with the simplest structure and chemical 
formula so far. Its crystal structure naturally removes the issue 
on Ga/Mg disorder proposed for YbMgGaO$_4$. The diversity of the 
large family makes it an ideal playground for studying the QSL 
physics and exploring its promising applications. 

\begin{figure*}[t]
\includegraphics[width=16cm]{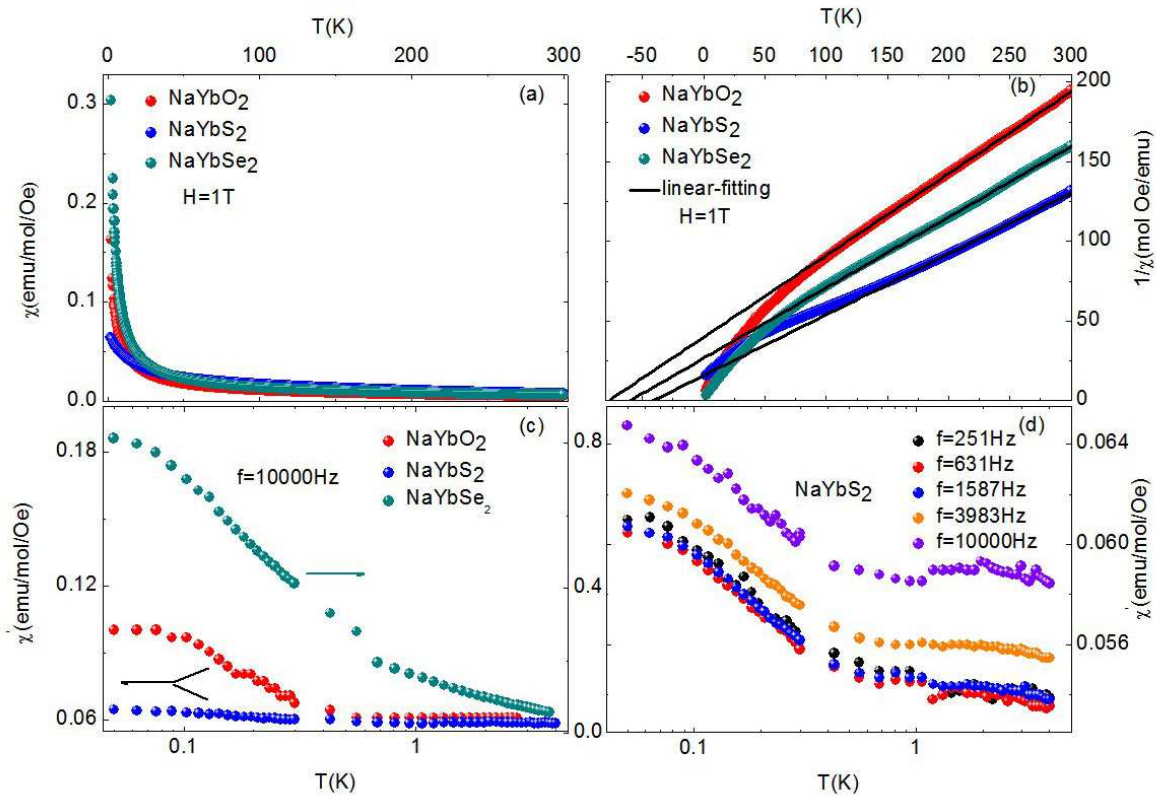}
\caption{(Color online.)  The DC (a \& b) and AC (c \& d) magnetic 
susceptibilities of polycrystalline NaYbCh$_2$ (Ch = O, S, Se).}
\label{fig2}
\end{figure*}

\emph{Sample preparation and experimental methods.}---The polycrystals of 
NaReO$_2$ (Re = Yb, Lu) were synthesized by the method of solid-state 
reaction under high temperatures:
\begin{equation}
\text{Na}_2\text{CO}_3 + \text{Re}_2\text{O}_3 \rightarrow 
2\text{NaReO}_2 + \text{O}_2 \quad	(\text{Re=Yb, Lu}).
\end{equation}
Na$_2$CO$_3$ and Re$_2$O$_3$ powders were mixed in a dry process 
(mixing molar ratio: Na$_2$O : Yb$_2$O$_3$ = 2.5 : 1) and shaped into 
a pellet by isostatic pressing (50MPa, 2min). Shaped samples were 
heated at 900 degrees for 9 hours. After the heating, the samples 
were ground and washed with distilled water and ethanol, at lastly 
dried in air at room temperature for 48 hours. The polycrystals of 
NaReS$_2$ (Re = La, Ce, Pr, Nd, Sm, Eu, Gd, Tb, Dy, Ho, Er, Tm, Yb, Lu) 
were synthesized by the method of solid-state reaction under high 
temperature:
\begin{equation}
\text{Na}_2\text{S} + 2\text{Re} + 3\text{S} \rightarrow 2\text{NaReS}_2,
\end{equation}
where Re = Lu, Se, Tm, Er, Ho, Dy, Tb, Gd, Eu, Sm, Nd, Pr, Ce, La. 
The Na$_2$S, Re and S powder was mixed in Ar environment glove box. 
The mixed powders were placed in a graphite crucible and vacuum 
packaged with quartz tube. Packaged samples were heated at 850 degrees 
for 48 hours. After the heating, the sample were powder and washed 
with distilled water and then dried in air at 50 degrees for 6 hours. 
The synthesis of polycrystals of NaReSe$_2$ (Re = Er, Yb, Lu) were 
similar with the synthesis of powders of NaReS$_2$. The temperature 
was adjusted to 900 degrees.

We have also successfully grown high quality NaYbSe$_2$ single crystals. 
The growth conditions of single crystals are more rigorous than NaYbSe$_2$ 
polycrystals. Na$_2$Se, Yb and Se powders were mixed in Ar environment 
glove box (mixing molar ratio: Na$_2$Se : Yb : Se = 1 : 2 : 12). 
The mixed powder was placed in special quartz tube that can withstand 
higher pressure. Packaged samples were headed at 1000 degrees for 48 hours. 
After the heating, we can observe 2-3mm size single crystals. We made a simple
resistance check for the crystals and the resistance is overranged and thus NaYbSe$_2$
is confirmed to be a good insulator.

Powder XRD profiles were measured with Bruker-D8 by step scanning. The TOPAS 
program was used for Rietveld crystal structure refinements. The temperature 
dependence of magnetic susceptibility from 1.8K to 300K was measured with a 
SQUID magnetometer (Quantum Design Magnetic Property Measurement System, MPMS) 
under both ZFC and FC for all the samples with the brass sample holder. 
The AC susceptibility measurements from 50mK to 4K were performed using 
a dilution refrigeration system (DR). The polycrystalline sample was
pressed into a thin plate and fixed on a sample holder with GE vanish. 
The heat capacity measurements from 2K to 30K were performed using PPMS 
(Quantum Design Physical Property Measurement System) and DR was employed 
for the measurements from 50mK to 4K. The plate sample was mounted on a 
sample holder with N grease for a better thermal contact.

\begin{table}
\begin{tabular}{lcccc}
\hline\hline
Ch &  Space group & C & $\Theta_{\text{CW}}$/K & $\mu_{\text{eff}}$ ($\mu_{\text B}$) \\
\hline\hline
O & R$\bar{3}$m & 1.9448 & $-77.35$ & 3.94 
\\
S & R$\bar{3}$m & 2.6166 & $-40.98$ & 4.57 
\\
Se& R$\bar{3}$m & 2.2550 & $-59.84$ & 4.24 
\\
\hline\hline
\end{tabular}
\caption{Parameters extracted from Curie-Weiss fitting for NaYbCh$_2$ (Ch= O, S, Se). }
\label{tab1}
\end{table}

\begin{figure*}[t]
\includegraphics[width=16cm]{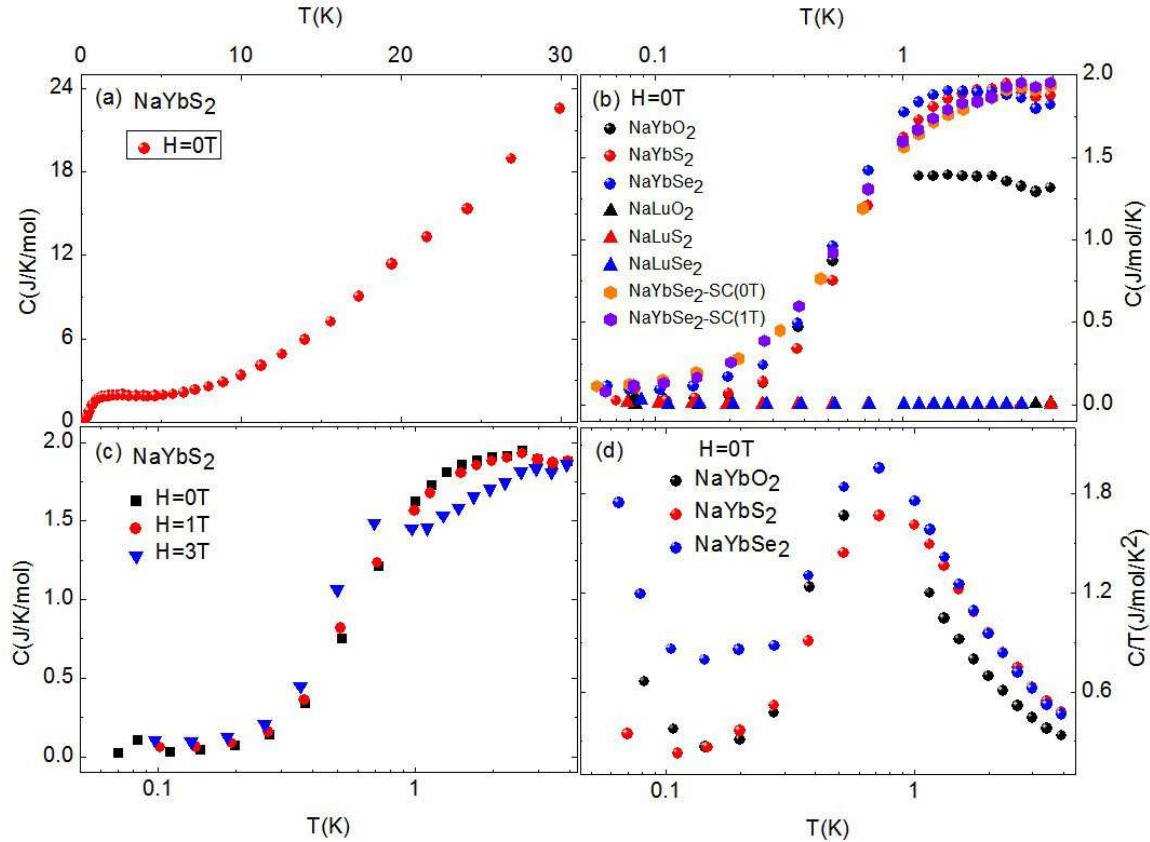}
\caption{(Color online.) Specific heat measurements on NaYbCh$_2$ (Ch = O, S, Se). 
Single crystal (SC) NaYbSe$_2$ and polycrystalline NaYbO$_2$ and NaYbS$_2$ were used in 
the measurements.}
\label{fig3}
\end{figure*}

\emph{Results and discussions.}---In Fig.~\ref{fig1}, we depict the crystal structure 
of the rare-earth chalcogenides and the Rietveld refinements for three representative 
samples NaYbO$_2$, NaYbS$_2$ and NaYbSe$_2$, where the detailed structural information 
extracted from the refinements can be found in the Supplementary Materials. 
The system has an R$\bar{3}$m space group symmetry, and the magnetic ions 
form flat triangular layers that are well separated. The ReCh$_6$ octahedra 
are connected in an edge-sharing fashion. The local crystal-field environment 
around the magnetic ions is exactly analogous to the case of YbMgGaO$_4$. 
Therefore, one expects a similar crystal-field splitting scheme of the 
Yb$^{3+}$ ions as the one in YbMgGaO$_4$. This means that a spin-orbital-entangled 
effective spin ${S=1/2}$ local moment should hold for our case. 
Similar to YbMgGaO$_4$, the anti-symmetric Dzyaloshinskii-Moriya 
interaction is prohibited by the inversion symmetry of the system.

The Mg/Ga disorder in the non-magnetic layers of YbMgGaO$4$ has been extensively 
discussed and is still under debate. Whether or how much this non-magnetic disorder 
impacts on the Yb magnetic properties is unclear in this stage. In some experiments 
and theoretical calculations, the disorder was considered to play a dominant role 
in contributing to the low-energy excitations. As a comparison, there is no such 
disorder in this family of rare-earth chalcogenides, due to the structural simplicity. 
The issue on disorder is completely removed for this family of materials. We further 
made the analysis of element ratio (See Supplementary Materials), which is close to 
the nominal ratio. This rules out the possibility of the disorder caused by element 
deficiency. If one still concerns about the active monovalent ions like Na$^+$ 
and K$^+$, he will have plenty of choices of the heavy monovalent ones such as 
Rb$^+$, Cs$^+$, Cu$^+$ and Ag$^+$, etc.

For the selected sub-family NaYbCh$_2$ (Ch = O, S, Se), we measured the DC magnetic 
susceptibility in the range of 2-300K and the AC susceptibility from 50mK to 4K. 
The results are presented in Fig.~\ref{fig2}. The Curie-Weiss fitting was made 
from 150K to 300K according to the crystal-field splitting in YbMgGaO$_4$ and 
the fitting results are summarized in Table.~\ref{tab1}. 
Considering the small interaction energy scale of rare-earth moments, this fitting range
may not be quite sufficient to characterize the low-energy magnetic physics of the system.
A lower fitting range may be required in the future. Nevertheless, the negative Curie-Weiss
temperatures suggest an antiferromagnetic coupling in all the samples. 
Excitingly, the Curie-Weiss temperatures are much larger than that of YbMgGaO$_4$ 
because of the smaller distances between nearest-neighbor Yb$^{3+}$ in 
the rare-earth chalcogenides. The AC susceptibility for the three samples   
shows no sign of long-range magnetic ordering. The measurements under 
various frequencies further confirms that there is no spin freezing either. 
Interestingly, the susceptibility saturation in the zero temperature limit
is clearly observed for all the three samples. This should be one of the
consequences caused by strong spin-orbit coupling rather than a sign of 
finite density of spin excitations. Under the strong spin-orbit coupling,
the total magnetization is not a good quantum number and cannot be used to 
label the many-body eigenstate. The many-body eigenstate would be a mixture 
of states with different total magnetizations. The magnetic susceptibility 
would always be a constant. We also find that the low-temperature 
susceptibility of NaYbSe$_2$ is obviously larger than that of the 
other two compounds. In fact, the distance between nearest neighbor
Yb$^{3+}$ ions in NaYbSe$_2$ is larger, while its Curie-Weiss 
temperatures and the moment obtained from the Curie-Weiss fitting 
look comparable to the other two.

\begin{figure}[t]
\includegraphics[width=8.5cm]{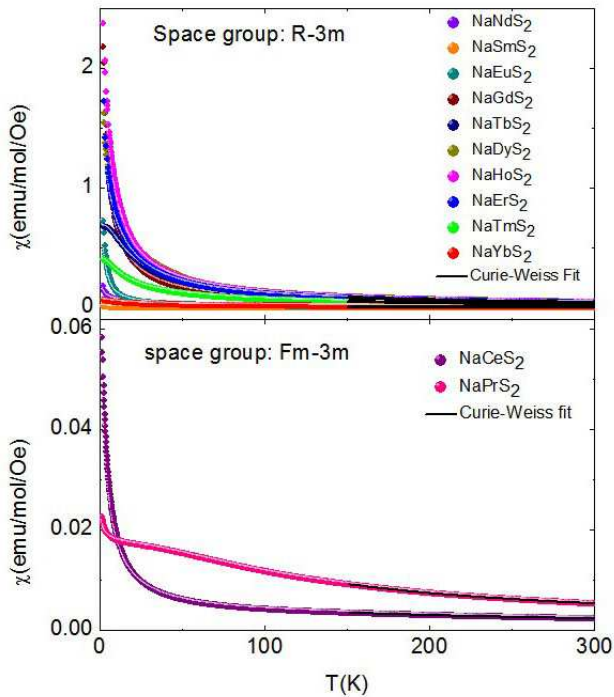}
\caption{(Color online.) Magnetic susceptibility of NaReS$_2$ (Re = Ce-Yb) 
in the range of 2K to 300K.} 
\label{fig4}
\end{figure}

Our specific heat results are shown in Fig.~\ref{fig3}. There is no 
obvious transition down to 50mK in these compounds, and it is consistent 
with the conclusion from the magnetic susceptibility data. We observed no 
apparent change or shift with applying a magnetic field up to 3T 
(see Fig.~\ref{fig3}c). On the other hand, an upturn is observed below 
100mK in the $C/T$-$T$ plot (see Fig.~\ref{fig3}d). This may arise from
the nuclear Schottky anomaly due to the nuclear spins. The upturn makes 
it difficult to obtain the intrinsic trend of the specific heat below 100mK 
and to conclude whether the system is a gapless or gapped QSL. Thus, more 
detailed magnetic and thermodynamic experiments are required in the future.
The broad peak around 1K in the $C/T$-$T$ plot is considered to be a consequence
of preserving entropy.
Here we point out that there is no obvious disorder in the present case 
and we still do not observe any long-range magnetic ordering or freezing 
that points to a possible QSL ground state. This means that the rare-earth 
triangular system, including rare-earth chalcogenides reported here and 
YbMgGaO$_4$, intrinsically hosts the QSL state that is not stabilized by or 
even originated from the Ga/Mg charge disorder.

The measurements discussed above are based on the sub-family NaYbCh$_2$ 
(Ch = O, S, Se). Towards a comprehensive view of the large family, we fixed 
Na and S, and systematically synthesized the other sub-family NaReS$_2$ 
(Re = La - Lu). The Rietveld refinements for all the fourteen compounds have been 
made and the detailed structural parameters can be found in the Supplementary Materials. 
The family members from Nd to Lu preserve the high lattice symmetry of R$\bar{3}$m. 
But the three end members, NaLaS$_2$, NaCeS$_2$ and NaPrS$_2$ show a cubic 
structure with the space group symmetry of Fm$\bar{3}$m. Clearly this is   
caused by the larger ionic radii of La, Ce and Pr. The magnetic susceptibility 
measurements from 2K to 300K have been carried out on the sub-family and   
the results are presented in Fig.~\ref{fig4}. No obvious magnetic transition 
is observed in all the compounds with the R$\bar{3}$m symmetry. Similar to 
the R$\bar{3}$m brothers, the cubic NaCeS$_2$ also shows no sign of magnetic 
transition from 2K to 300K, as the perfect triangular lattice formed by 
the magnetic ions remains undistorted and the strong geometrical frustration 
is always there. In contrast, an anomaly appears around 10K in the susceptibility 
of NaPrS$_2$ (Fm$\bar{3}$m).

The Curie-Weiss fitting was performed from 150K to 300K, assuming a reasonably 
high crystal field splitting. The full results are presented in Table~\ref{tab2}. 
We can see that the exchange couplings vary from sample to sample. In the other 
words, we have the opportunity to select the compounds with various exchange coupling. 
Beyond this, one can further tune the charge gaps of the family members by element 
substitution. The absorption spectra (see Supplementary Materials) indicate that the 
charge gaps are roughly 4.5eV, 2.7eV and 1.9eV for NaYbO$_2$, NaYbS$_2$, and NaYbSe$_2$, 
respectively. The variable and small charge gaps may allow the system to access   
a Mott-metal transition by applying doping or pressures. Such a possibility opens 
up the interesting direction of Mott transitions out of a QSL~\cite{PhysRevB.78.045109}.
This transition was argued to be continuous by noticing that the Landau damping term
scales like a mass term for the bosonic charge and then identifying the transition
as an usual superfluid-Mott transition~\cite{PhysRevB.78.045109}.  
Thus, these exciting advantages stem from the rich diversity of the family. 
In fact, we have made a careful literature research and found that most of the 
family members have the high-symmetry of R$\bar{3}$m and hence are potential QSL
materials (see Supplementary Materials). This suggests that the family is an ideal 
playground, on which we can tune the basic material parameters or exchange coupling 
to explore the QSL physics and develop its possible applications.

\begin{table}[t]
\begin{tabular}{lcclll}
\hline\hline 
Re & Space group & C & $\Theta_{\text{CW}}$/K & $\mu_{\text{eff}}$ (Obs.) & $\mu_{\text{eff}}$ (Cal.) \\
\hline\hline 
Ce & F$m\bar{3}m$ & 1.1145 & $-164.42$ & 2.99$\mu_{\text{B}}$ & 2.54$\mu_{\text{B}}$ 
\\
Pr & F$m\bar{3}m$ & 1.91757 & $-57.48$ & 3.92$\mu_{\text{B}}$ & 3.58$\mu_{\text{B}}$ 
\\
Nd & R$\bar{3}$m & 1.78198 & $-25.35$ & 3.77$\mu_{\text{B}}$ & 3.62$\mu_{\text{B}}$ 
\\
Sm & R$\bar{3}$m & 0.71761 & $-343.05$ & 2.40$\mu_{\text{B}}$ & 0.84$\mu_{\text{B}}$
\\
Eu & R$\bar{3}$m & 3.19082 & $-106.13$ & 5.05$\mu_{\text{B}}$ & 3.6$\mu_{\text{B}}$ 
\\
Gd & R$\bar{3}$m & 8.5521 & $-1.98$ & 8.27$\mu_{\text{B}}$ & 7.94$\mu_{\text{B}}$ 
\\ 
Tb & R$\bar{3}$m & 12.87953 & $-9.49$ & 10.15$\mu_{\text{B}}$ & 9.72$\mu_{\text{B}}$
\\
Dy & R$\bar{3}$m & 14.55906 & $-9.39$ & 10.79$\mu_{\text{B}}$ & 10.63$\mu_{\text{B}}$ 
\\
Ho & R$\bar{3}$m & 13.8468 & $-5.90$ & 10.52$\mu_{\text{B}}$ & 10.60$\mu_{\text{B}}$
\\
Er & R$\bar{3}$m & 12.13713 & $-4.62$ & 9.85$\mu_{\text{B}}$ & 9.59$\mu_{\text{B}}$
\\
Tm & R$\bar{3}$m & 7.40589 & $-3.83$ & 7.69$\mu_{\text{B}}$ & 7.57$\mu_{\text{B}}$
\\
Yb & R$\bar{3}$m & 2.90963 & $-63.74$ & 4.82$\mu_{\text{B}}$ & 4.54$\mu_{\text{B}}$
\\
\hline\hline
\end{tabular}
\caption{Parameters extracted from the Curie-Weiss fitting for NaReS$_2$ (Re = Ce-Yb).}
\label{tab2}
\end{table}

\emph{Summary.}---In summary, we have synthesized rare-earth chalcogenides AReCh$_2$ that exhibit 
a delafossite structure, and made structural and thermodynamics characterizations. 
The family has a lattice symmetry R$\bar{3}$m, and the magnetic ions are antiferromagnetically 
coupled and form perfect triangular layers. The magnetic susceptibility and specific heat 
measurements down to 50mK indicate no sign of long-range magnetic ordering or transition. 
The family removes the disorder issue raised in YbMgGaO$_4$ and also suggests the QSL 
physics is probably not from disorder. The unique advantages, such as various charge gaps and 
exchange coupling, suggest that the family may be an ideal platform for the further study of QSLs.

\emph{Acknowledgments.}---We thank Zicheng Wen for assisting the absorption 
measurements and Feng Jin for organizing the references. This work is supported by the Ministry of Science and Technology of 
China (2016YFA0300504 \& 2017YFA0302904 \& 2016YFA0301001) and the NSF of China 
(11774419, 11474357, 11822412, 11774423 \& 11574394).

Note added: Upon the completion of this work, we become aware of Ref.~\onlinecite{NaYbS2} 
that focused on NaYbS$_2$ and proposed it as a spin liquid candidate.

\bibliography{refs}

\end{document}